\documentclass[aps,preprint,onecolumn,11pt,showkeys]{revtex4}%
\usepackage{amsfonts}
\usepackage{amsmath}
\usepackage{amssymb}
\usepackage{graphicx}
\usepackage[ansinew]{inputenc}
\usepackage[usenames,dvipsnames]{pstricks}
\usepackage{subfigure}
\usepackage{pdfpages}
\usepackage{epsfig}
\usepackage{pst-grad}
\usepackage{pst-plot}
\usepackage[colorlinks,hyperindex]{hyperref}%
\providecommand{\U}[1]{\protect\rule{.1in}{.1in}}
\hypersetup{
colorlinks,
citecolor=blue,
linkcolor=black,
urlcolor=black,
}

\begin{document}
\title{Analytical general solutions for static wormholes in $f(R,T)$ gravity}
\author{P.H.R.S. Moraes, R.A.C. Correa and R.V. Lobato$^{}\medskip$\thanks{moraes.phrs@gmail.com}}
\affiliation{{\small {$^{}$ Instituto Tecnológico de Aeronáutica, 12228-900, São José dos
Campos, SP, Brazil}\medskip}}
%\author{$^{}\medskip$\thanks{rafael.couceiro@ufabc.edu.br}}
%\affiliation{{\small {$^{}$ Instituto Tecnológico de Aeronáutica, 12228-900, São José dos
%Campos, SP, Brazil}\medskip}\\
%{\small {$^{}$ Universidade Estadual Paulista, 12516-410, Guaratinguetá, SP,
 %     Brazil}\medskip}}
%\author{$^{}\medskip$\thanks{rvlobato@ita.br}}
%\affiliation{{\small {$^{}$ Instituto Tecnológico de Aeronáutica, 12228-900, São José dos
%Campos, SP, Brazil}\medskip}}
\begin{abstract}
Originally proposed as a tool for teaching the general theory of relativity,
wormholes are today approached in many different ways and are seeing as an
efficient alternative for interstellar and time travel. Attempts to achieve
observational signatures of wormholes have been growing as the subject has
became more and more popular. In this article we investigate some $f(R,T)$
theoretical predictions for static wormholes, i.e., wormholes whose throat
radius can be considered a constant. Since the $T$-dependence in $f(R,T)$
gravity is due to the consideration of quantum effects, a further
investigation of wormholes in such a theory is well motivated. We obtain the energy conditions of static wormholes in $f(R,T)$ gravity and apply an analytical approach to find the solutions. We highlight that our results are in agreement with
previous solutions presented in the literature.

\end{abstract}
\maketitle

\section{Introduction}

\label{sec:int}

Wormholes (WHs) are known to provide a conceivable method for rapid
interstellar travel. They connect two distant regions of the Universe and are
a solution for the Einstein's field equations of General Relativity (GR). The
structure of a WH is tube-like, asymptotically flat from both sides. Depending
on its theoretical construction, the radius of the WH's throat can be
considered either constant, namely static WHs (SWHs), or variable, namely
non-static or cosmological WHs.

In GR, traversable WHs can exist only in the presence of exotic matter, that
violates the null energy condition (NEC) \cite{morris/1988}. On the other
hand, if alternative theories of gravity are capable of describing the WH
geometry, they achieve this without necessarily invoking exotic matter.
Such a description was made in $f(R)$ gravity theories
\cite{pavlovic/2015,lobo/2014,rahaman/2014,harko/2013,de_benedicts/2012,lobo/2009,furey/2005}%
. Other theories of gravity have also described the (non-exotic) matter
content and geometry of WHs, such as Mimetic gravity \cite{myrzakulov/2016},
Eddington-Inspired Born-Infeld gravity
\cite{harko/2015,shaikh/2015,eiroa/2012}, Dvali-Gabadadze-Porrati braneworld
model \cite{richarte/2013} and Gauss-Bonnet theory
\cite{mehdizadeh/2015,kanti/2011,maeda/2008}, among many others.

In 2011, T. Harko and collaborators proposed an extension of the $f(R)$
theories, by inserting in the gravitational action of the model not only a
general dependence on the Ricci scalar $R$, but also on the trace of the
energy-momentum tensor $T$, namely the $f(R,T)$ gravity \cite{harko/2011}.
Such an alternative gravity theory has been tested in areas such as Cosmology
\cite{mrc/2016,mc/2016,cm/2016,ms/2016,moraes/2016,moraes/2015,baffou/2015,shabani/2014,moraes/2014,shabani/2013}%
, Thermodynamics \cite{mmm/2016,harko/2014,sharif/2012,sharif/2012b},
Astrophysics of compact objects
\cite{mam/2016,zubair/2016,shamir/2015,noureen/2015,zubair/2015} and
gravitational waves \cite{amam/2016}. Despite these efforts, the information
content of WHs in $f(R,T)$ theories is still poor. In fact, a particular case
of the WH geometry, in which its redshift function $\varphi$ does not depend
on time nor on the spatial coordinates, was studied in
\cite{azizi/2013,zubair/2016b}. Since the dependence on $T$ in the $f(R,T)$
gravity arises as a consequence of the consideration of quantum effects, it
might be interesting to further investigate WHs in such a theory.

Although no observational evidences of WHs have been found so far, recently
some important contributions on this regard were proposed. For instance, the
existence of WHs in the galactic halo regions was discussed in
\cite{rahaman/2014b}. It was shown that the space-time of a galactic halo may
be described by a traversable WH geometry, fitting with its observed flat
galactic rotation curve. Further investigating such an issue, it was shown
that the detection of these WHs is a distinct possibility by means of
gravitational lensing \cite{kuhfittig/2014}. This possibility was analyzed
also in \cite{nandi/2006}.

It is known that supermassive black hole candidates at the center of most
galaxies might be WHs created in the early Universe. A method for
distinguishing between black holes and WHs with orbiting hot spots was
presented in \cite{li/2014} and by their Einstein-ring systems in
\cite{tsukamoto/2012}.

Specific signatures in the electromagnetic spectrum were obtained for thin
accretion disks surrounding static spherically symmetric WHs in
\cite{bambi/2013,harko/2009,harko/2008}, leading to the possibility of
distinguishing WH geometries by using astrophysical observations of such
emission spectra. Other proposals for finding WH astrophysical signatures can
be appreciated in \cite{safonova/2002,safonova/2001,torres/1998,cramer/1995}.

With continuous advances in what concerns WH detection attempts, it is worth
to collect further predictions of their matter-geometry content. That is the
purpose of the present article, in which such a collection will be made from
the $f(R,T)$ theory of gravity perspective.

The paper is organized as follows: in Section \ref{sec:frt} we present a brief
review of the $f(R,T)$ formalism. In Section \ref{sec:swhm}, we present the
Morris-Thorne metric, usually assigned for the geometry of WHs. We also show
some conditions that must be obeyed by the metric potentials of the WH
geometry. In Section \ref{sec:swhfrt}, we develop the field equations for the
SWH metric in the $f(R,T)$ gravity theory. We present the correspondent energy conditions and obtain solutions for the WH physical and geometrical parameters. Our
results are discussed in Section \ref{sec:dis}.

\section{The $f(R,T)$ gravity}

\label{sec:frt}

The $f(R,T)$ theory of gravity considers its gravitational action to be
dependent on a general function of both $R$ and $T$. The $T$-dependence is
motivated by the consideration of quantum effects (conformal anomaly). The
$f(R,T)$ action reads%

\begin{equation}
\label{frt1}\mathcal{S}=\int\left[  \frac{f(R,T)}{16\pi}+\mathcal{L}%
_{m}\right]  \sqrt{-g}d^{4}x.
\end{equation}
In (\ref{frt1}), $f(R,T)$ is the general function of $R$ and $T$,
$\mathcal{L}_{m}$ is the matter lagrangian density, $g$ is the determinant of
the metric $g_{\mu\nu}$ and the system of units here adopted is such that
$c=1=G$.

The matter lagrangian density above contains informations about the matter
fields of the model. Here, it will be assumed $\mathcal{L}_{m}=-\mathcal{P}$,
with $\mathcal{P}$ being the total pressure. Since the WH matter content is
being considered here, the energy-momentum tensor of an anisotropic fluid,
such as \cite{morris/1988}%

\begin{equation}
\label{frt2}T_{\mu\nu}=(\rho+p_{t})u_{\mu}u_{\nu}+p_{t}g_{\mu\nu}+(p_{r}%
-p_{t})\chi_{\mu}\chi_{\nu},
\end{equation}
will be taken into account. In (\ref{frt2}), $\rho$ is the matter-energy
density, $p_{r}$ and $p_{t}$ are, respectively, the radial pressure measured
in the direction of $\chi_{\mu}$ and transverse pressure measured in the
direction orthogonal to $\chi_{\mu}$, with $\chi_{\mu}$ being some space-like
vector orthogonal to the $4$-velocity $u_{\mu}$.

The field equations of the theory are obtained by varying the action with
respect to $g_{\mu\nu}$ and read%

\begin{equation}
\label{frt3}f_{R}(R,T)R_{\mu\nu}-\frac{1}{2}f(R,T)g_{\mu\nu}+(g_{\mu\nu}%
\Box-\nabla_{\mu}\nabla_{\nu})f_{R}(R,T)=8\pi T_{\mu\nu}+f_{T}(R,T)(T_{\mu\nu
}+\mathcal{P}g_{\mu\nu}).
\end{equation}
In (\ref{frt3}), $f_{R}(R,T)\equiv\partial f(R,T)/\partial R$, $R_{\mu\nu}$ is
the Ricci tensor, $\Box$ is the D'Alambert operator, $\nabla_{\mu}$ is the
covariant derivative, $f_{T}(R,T)\equiv\partial f(R,T)/\partial T$ and
$\mathcal{P}=(p_{r}+2p_{t})/3$ is the total pressure.

From (\ref{frt3}), the covariant derivative of the energy-momentum tensor reads%

\begin{equation}
\label{frt4}\nabla^{\mu}T_{\mu\nu}=-\frac{f_{T}(R,T)}{f_{T}(R,T)+8\pi}\left[
(T_{\mu\nu}+\mathcal{P}g_{\mu\nu})\nabla^{\mu}\ln f_{T}(R,T)+\frac{1}{2}%
g_{\mu\nu}\nabla^{\mu}(\rho-\mathcal{P})\right]  ,
\end{equation}
in which it was already considered $T=\rho-3\mathcal{P}$.

In order to construct exact WH solutions in $f(R,T)$ theory, it will be
necessary to consider a specific form for the function $f(R,T)$ to be
substituted in (\ref{frt3})-(\ref{frt4}). Here it will be assumed
$f(R,T)=R+2f(T)$, with $f(T)=\lambda T$ and $\lambda$ a constant. Such a functional form was
proposed by the $f(R,T)$ gravity authors themselves \cite{harko/2011} and
since then it has been broadly applied in $f(R,T)$ models. Moreover, it
benefits from the fact that GR is recovered by simply making $\lambda=0$. Such
an assumption yields, for (\ref{frt3})-(\ref{frt4}), the following%

\begin{equation}
\label{frt5}G_{\mu\nu}=8\pi T_{\mu\nu}+2\lambda[T_{\mu\nu}+(\rho
-\mathcal{P})g_{\mu\nu}],
\end{equation}
\begin{equation}
\label{frt6}\nabla^{\mu}T_{\mu\nu}=\left(  \frac{\lambda}{2\lambda+8\pi
}\right)  g_{\mu\nu}\nabla^{\mu}(\mathcal{P}-\rho),
\end{equation}
with $G_{\mu\nu}$ being the usual Einstein tensor.

\section{The static wormhole metric}

\label{sec:swhm}

The Morris-Thorne metric, which can describe the geometry of a SWH, is given
by \cite{morris/1988}%

\begin{equation}
\label{swhm1}ds^{2}=e^{2\varphi(r)}dt^{2}-\left[1-\frac{b(r)}{r}\right]^{-1}dr^{2}-r^{2}(d\theta^{2}+sin^{2}\theta d\phi^{2}).
\end{equation}
In (\ref{swhm1}), $\varphi(r)$ is the redshift function and $b(r)$ is the
shape function.

The radial coordinate in a WH needs to non-monotonically decrease from
infinity to a minimal value $r_{0}$ at the throat, where $b(r_{0})=r_{0}$, and
then increase to infinity.

Furthermore, the metric potential $\varphi(r)$ must be finite everywhere in
order to the WH be traversable. The other metric potential, $b(r)$, needs to
obey the following conditions \cite{morris/1988}%

\begin{equation}
\label{swhm2}1-\frac{b}{r}>0,
\end{equation}
\begin{equation}
\label{swhm3}\frac{b-b^{\prime}r}{b^{2}}>0,
\end{equation}
with primes denoting radial derivatives. Moreover, at the throat, the
condition $b^{\prime}(r_{0})<1$ is imposed in order to have WH solutions.

\section{Static wormholes in $f(R,T)$ gravity}

\label{sec:swhfrt}

In order to start constructing the $f(R,T)$ gravity SWH, let us begin by
developing Eq.(\ref{frt5}) for metric (\ref{swhm1}). We, then, have%

\begin{equation}
\label{swhfrt1}\frac{b^{\prime}}{r^{2}}=8\pi\rho+2\lambda\left(  2\rho
-\frac{p_{r}+2p_{t}}{3}\right)  ,
\end{equation}
\begin{equation}
\label{swhfrt2}\frac{1}{r}\left[  \frac{b}{r^{2}}+2\varphi^{\prime}\left(
\frac{b}{r}-1\right)  \right]  =-8\pi p_{r}+2\lambda\left[  \rho-\frac{2}%
{3}(2p_{r}+p_{t})\right]  ,
\end{equation}
\begin{equation}
\label{swhfrt3}\frac{1}{2r}\left[  \frac{1}{r}\left(  \varphi^{\prime
}b+b^{\prime}-\frac{b}{r}\right)  +2(\varphi^{\prime\prime}+\varphi^{\prime
2})b-\varphi^{\prime}(2-b^{\prime})\right]  -(\varphi^{\prime\prime}%
+\varphi^{\prime2})=-8\pi p_{t}+2\lambda\left(  \rho-\frac{p_{r}+5p_{t}}%
{3}\right)  .
\end{equation}

Also, from Eqs.(\ref{frt6})-(\ref{swhm1}), one can show that

\begin{equation}
p_{r}^{\prime}+\varphi^{\prime}(\rho+p_{r})=\frac{\lambda}{2\lambda+8\pi
}\left(  \rho^{\prime}-\frac{p_{r}^{\prime}+2p_{t}^{\prime}}{3}\right)  .
\label{swhfrt4}%
\end{equation}

The advantage of constructing WHs in an alternative gravity theory, such as the $f(R,T)$ gravity, is the fact that the extra terms of its field equations (when compared to GR field equations) can provide the WH obedience of the energy conditions. In other words, with the presence of the extra terms, it is unnecessary to invoke exotic fluids to permeate the WH in a modified theory of gravity, departing from the GR case. Below we will investigate the $f(R,T)$ gravity WHs under some energy conditions.

Let us start by analysing the $f(R,T)$ SWH from the NEC perspective. The NEC reads $T_{\mu\nu}^{eff}u_\mu u_\nu\geq0$ \cite{morris/1988}, with $T_{\mu\nu}^{eff}$ the effective energy-momentum tensor and $u_\mu$ being some null vector. This can be rewritten as $\rho_{eff}+p_{eff}\geq0$. In \cite{alvarenga/2013}, some $f(R,T)$ models were tested from energy conditions. It was shown that in $f(R,T)$ gravity, 

\begin{equation}\label{ec1}
\rho_{eff}=\rho+f(T)+2\rho f_T(T),
\end{equation}
\begin{equation}\label{ec2}
p_{eff}=-f(T),
\end{equation}
with $f_T(T)\equiv df(T)/dT$. From (\ref{ec1})-(\ref{ec2}), the NEC yields $\lambda\leq-1/2$, which is in accordance with some previous calculations \cite{azizi/2013}.

Before checking the weak energy condition (WEC) of WHs in $f(R,T)$ gravity, we note that Eqs.(\ref{swhfrt1})-(\ref{swhfrt4}) have a non-linear character. Besides that, they are coupled equations. It might be important to
remark that the presence of nonlinearity is not surprising, given that such a
behavior is found in a wide range of areas of Physics \cite{correa1, correa2,
correa3, correa5, correa7, correa9, correa11, correa12, correa13}. Because of
the nonlinearity, we are led to ask if the problem can be analytically
resolved. We show below that, indeed, it is possible to obtain an interesting
class of analytical solutions for the material quantities and geometrical
parameters of the $f(R,T)$ gravity SWH.

Thus, in order to find analytical solutions, it is both useful and natural to use
an equation of state (EoS), i.e., a relation between pressure and matter-energy density. Here, we will apply the following EoS 
\begin{align}
p_{r} &  =\alpha_{r}+A_{r}\rho,\label{q1}\\
p_{t} &  =\beta_{t}+B_{t}\rho,\label{q2}%
\end{align}
\noindent where\noindent\ $\alpha_{r}$, $A_{r}$, $\beta_{t}$ and $B_{t}$ are
constants, which will be further defined in the text. The same sort of EoS was already applied to another WH studies, such as \cite{azizi/2013,jamil/2013,lobo/2009,zubair/2016b}, among others. Note that $\alpha_{r}$ and $\beta_{t}$ can be viewed as specifying translation effects, while $A_{r}$ and $B_{r}$ are related to dilation effects. The translation in both radial and transverse pressure structures is a symmetry operation, in the sense that the structures
remain the same when observed from a starting point or from this point
shifted by the translation constants. Moreover, the dilation effects can give
rise to some kind of amplification or compactification in the pressure
structures.

Now, using Eqs.(\ref{q1})-(\ref{q2}) into Eq.(\ref{swhfrt1}), we get%
\begin{equation}
\frac{b^{\prime}}{r^{2}}=\left[   8\pi+4\lambda  -\frac
{2\lambda}{3}\left(  A_{r}+2B_{t}\right)  \right]  \rho-\frac{2\lambda}%
{3}\left(  \alpha_{r}+2\beta_{t}\right)  . \label{q3}%
\end{equation}

In order to eliminate the dependence on $\rho$ in the equation above, we
impose that%
\begin{equation}
8\pi+4\lambda-\frac{2\lambda}{3}\left(  A_{r}+2B_{t}\right)  =0 \label{q4}%
\end{equation}
and obtain the following solution for Eq.(\ref{q3})%
\begin{equation}
b(r)=b_{0}-c_{0}r^{3}, \label{q5}%
\end{equation}
\noindent where $b_{0}$ is an arbitrary constant of integration, which for the
sake of simplicity we take as null. Moreover, we are using the definition%
\begin{equation}
c_{0}\equiv\frac{2\lambda}{9}\left(  \alpha_{r}+2\beta_{t}\right)  .
\label{q6}%
\end{equation}

In possession of solution (\ref{q5}) and making use of Eqs.(\ref{q1}%
)-(\ref{q2}), we can rewrite Eq.(\ref{swhfrt2}) in the form%
\begin{equation}
2c_{0}\varphi^{\prime}(r^{2}-r)=c_{0}-\left[  8\left(  \pi+\frac{\lambda}%
{3}\right)  \alpha_{r}+\frac{4\lambda}{3}\beta_{t}\right]  -\left[  8\left(
\pi+\frac{\lambda}{3}\right)  A_{r}-2\lambda+\frac{4\lambda}{3}B_{t}\right]
\rho. \label{q7.1}%
\end{equation}
Again, we will remove the dependence on $\rho$. This is achieved by setting%
\begin{equation}
8\left(  \pi+\frac{\lambda}{3}\right)  A_{r}-2\lambda+\frac{4\lambda}{3}%
B_{t}=0. \label{q8}%
\end{equation}

Consequently, we obtain the corresponding solution for the redshift function%
\begin{equation}
\varphi(r)=\varphi_{0}+\Gamma_{0}\ln\left[  \frac{R_{0}\left(  r-1\right)
}{r}\right]  , \label{q9}%
\end{equation}
\noindent where $\varphi_{0}$ is an arbitrary constant of integration and
$R_{0}=r_{0}/(r_{0}-1)$. Furthermore, $\Gamma_{0}$ is defined as%
\begin{equation}
\Gamma_{0}\equiv\frac{1}{2}-\frac{2}{c_{0}}\left[  2\left(  \pi+\frac{\lambda
}{3}\right)  \alpha_{r}+\frac{\lambda}{3}\beta_{t}\right]  . \label{q10}%
\end{equation}

At this point, it is important to remark that by choosing $\alpha_{r}=0$ and
$\beta_{t}=0$ we can recover some particular cases obtained in the literature,
where the redshift function $\varphi(r)=\varphi_{0}=$ \textit{constant}.

Note that the Eqs.(\ref{q4}) and (\ref{q8}) define a linear system with two
unknowns: $A_{r}$ and $B_{t}$. From such a system of equations, we obtain the
following solutions%
\begin{align}
\tilde{A}_{r}  &  =6\left(  2\pi+\lambda\right)  +\frac{\lambda^{2}-8\left(
\lambda+3\pi\right)  \left(  \lambda+2\pi\right)  }{3\pi},\label{q11}\\
\tilde{B}_{t}  &  =\frac{8\left(  3\pi+\lambda\right)  \left(  2\pi
+\lambda\right)  -\lambda^{2}}{6\pi}. \label{q12}%
\end{align}

\noindent where $\tilde{A}_{r}\equiv\lambda A_{r}$ and
$\tilde{B}_{t}\equiv\lambda B_{t}.$

To find the matter-energy density $\rho$ of the SWH, we insert the solutions
for $b(r)$ and $\varphi(r)$ and also the relations (\ref{q1}) and (\ref{q2})
into Eq.(\ref{swhfrt3}). After straightforward mathematical manipulations we find%

\begin{equation}
\rho(r)=\frac{\left[  3(c_{0}-8\pi\beta_{t})-2\lambda(\alpha_{r}+5\beta
_{t})\right]  r^{2}(r-1)^{2}-3\Gamma_{0}r(1+c_{0}r)+3\Gamma_{0}^{2}%
(1+c_{0}r)^{2}}{2r^{2}(r^{2}-1)^{2}\left[  12\pi B_{t}+\lambda\left(
A_{r}+5B_{t}-3\right)  \right]  }. \label{q13}%
\end{equation}

Now, let us use Eq.(\ref{swhfrt4}) to determine the constraints between $\alpha_{r}$
and $\beta_{t}$. By applying Eqs.(\ref{q5}), (\ref{q9}) and (\ref{q13}) into
Eq.(\ref{swhfrt4}), we get%
\begin{equation}
\sum\limits_{j=0}^{4}\mathcal{H}_{j}(\alpha_{r},\beta_{t})r^{n}=0,\label{q14}%
\end{equation}
\noindent where $\mathcal{H}_{j}(\alpha_{r},\beta_{t})$ are coefficients given
by%
\begin{equation}
\left.  \mathcal{H}_{0}(\alpha_{r},\beta_{t})=48A_{r}\pi\text{$\Gamma$}%
_{0}^{2}+24\pi\text{$\Gamma$}_{0}^{3}+24A_{r}\pi\text{$\Gamma$}_{0}%
^{3}+14A_{r}\text{$\Gamma$}_{0}^{2}\lambda+4B_{t}\text{$\Gamma$}_{0}%
^{2}\lambda+6\text{$\Gamma$}_{0}^{3}\lambda+6A_{r}\text{$\Gamma$}_{0}%
^{3}\lambda,\right.
\end{equation}%
\begin{equation}
\left.  \mathcal{H}_{1}(\alpha_{r},\beta_{t})=-24A_{r}\pi\text{$\Gamma$}%
_{0}-24\pi\text{$\Gamma$}_{0}^{2}-120A_{r}\pi\text{$\Gamma$}_{0}^{2}%
-7A_{r}\text{$\Gamma$}_{0}\lambda-2B_{t}\text{$\Gamma$}_{0}\lambda
-6\text{$\Gamma$}_{0}^{2}\lambda-34A_{r}\text{$\Gamma$}_{0}^{2}\lambda
-8B_{t}\text{$\Gamma$}_{0}^{2}\lambda,\right.
\end{equation}%
\begin{align}
&  \left.  \mathcal{H}_{2}(\alpha_{r},\beta_{t})=72A_{r}\pi\text{$\Gamma$}%
_{0}+24c_{0}\pi\text{$\Gamma$}_{0}+24A_{r}c_{0}\pi\text{$\Gamma$}_{0}%
+192B_{t}\pi^{2}\text{$\alpha$}_{r}\text{$\Gamma$}_{0}-192\pi^{2}\text{$\beta
$}_{t}\text{$\Gamma$}_{0}-192A_{r}\pi^{2}\text{$\beta$}_{t}\text{$\Gamma$}%
_{0}\right.  \nonumber\\
&  \left.  -24c_{0}\pi\text{$\Gamma$}_{0}^{2}-24A_{r}c_{0}\pi\text{$\Gamma$%
}_{0}^{2}+24c_{0}\pi\text{$\Gamma$}_{0}^{3}+24A_{r}c_{0}\pi\text{$\Gamma$}%
_{0}^{3}+21A_{r}\text{$\Gamma$}_{0}\lambda+6B_{t}\text{$\Gamma$}_{0}%
\lambda+6c_{0}\text{$\Gamma$}_{0}\lambda+6A_{r}c_{0}\text{$\Gamma$}_{0}%
\lambda\right.  \nonumber\\
&  \left.  -64\pi\text{$\alpha$}_{r}\text{$\Gamma$}_{0}\lambda+128B_{t}%
\pi\text{$\alpha$}_{r}\text{$\Gamma$}_{0}\lambda-128\pi\text{$\beta$}%
_{t}\text{$\Gamma$}_{0}\lambda-128A_{r}\pi\text{$\beta$}_{t}\text{$\Gamma$%
}_{0}\lambda-6c_{0}\text{$\Gamma$}_{0}^{2}\lambda-6A_{r}c_{0}\text{$\Gamma$%
}_{0}^{2}\lambda+6c_{0}\text{$\Gamma$}_{0}^{3}\lambda\right.  \nonumber\\
&  \left.  +6A_{r}c_{0}\text{$\Gamma$}_{0}^{3}\lambda-16\text{$\alpha$}%
_{r}\text{$\Gamma$}_{0}\lambda^{2}+20B_{t}\text{$\alpha$}_{r}\text{$\Gamma$%
}_{0}\lambda^{2}-20\text{$\beta$}_{t}\text{$\Gamma$}_{0}\lambda^{2}%
-20A_{r}\text{$\beta$}_{t}\text{$\Gamma$}_{0}\lambda^{2},\right.
\end{align}%
\begin{align}
&  \left.  \mathcal{H}_{3}(\alpha_{r},\beta_{t})=-48c_{0}\pi\text{$\Gamma$%
}_{0}-384B_{t}\pi^{2}\text{$\alpha$}_{r}\text{$\Gamma$}_{0}+384\pi
^{2}\text{$\beta$}_{t}\text{$\Gamma$}_{0}+384A_{r}\pi^{2}\text{$\beta$%
}t\text{$\Gamma$}_{0}-48A_{r}c_{0}\pi\text{$\Gamma$}_{0}^{2}-12c_{0}%
\text{$\Gamma$}_{0}\lambda\right.  \nonumber\\
&  \left.  +2A_{r}c_{0}\text{$\Gamma$}_{0}\lambda+4B_{t}c_{0}\text{$\Gamma$%
}_{0}\lambda+128\pi\text{$\alpha$}_{r}\text{$\Gamma$}_{0}\lambda-256B_{t}%
\pi\text{$\alpha$}_{r}\text{$\Gamma$}_{0}\lambda+256\pi\text{$\beta$}%
_{t}\text{$\Gamma$}_{0}\lambda+256A_{r}\pi\text{$\beta$}_{t}\text{$\Gamma$%
}_{0}\lambda-14A_{r}c_{0}\text{$\Gamma$}_{0}^{2}\lambda\right.  \nonumber\\
&  \left.  -4B_{t}c_{0}\text{$\Gamma$}_{0}^{2}\lambda+32\text{$\alpha$}%
_{r}\text{$\Gamma$}_{0}\lambda^{2}-40B_{t}\text{$\alpha$}_{r}\text{$\Gamma$%
}_{0}\lambda^{2}+40\text{$\beta$}_{t}\text{$\Gamma$}_{0}\lambda^{2}%
+40A_{r}\text{$\beta$}_{t}\text{$\Gamma$}_{0}\lambda^{2},\right.
\end{align}

\begin{align}
& \left.  \mathcal{H}_{4}(\alpha_{r},\beta_{t})=24c_{0}\pi\text{$\Gamma$}%
_{0}+24A_{r}c_{0}\pi\text{$\Gamma$}_{0}+192B_{t}\pi^{2}\text{$\alpha$}%
_{r}\text{$\Gamma$}_{0}-192\pi^{2}\text{$\beta$}_{t}\text{$\Gamma$}%
{0}-192A_{r}\pi^{2}\text{$\beta$}_{t}\text{$\Gamma$}_{0}+6c_{0}\text{$\Gamma
$}_{0}\lambda+6A_{r}c_{0}\text{$\Gamma$}_{0}\lambda\right.  \nonumber\\
& \left.  -64\pi\text{$\alpha$}_{r}\text{$\Gamma$}_{0}\lambda+128B_{t}%
\pi\text{$\alpha$}_{r}\text{$\Gamma$}_{0}\lambda-128\pi\text{$\beta$}%
_{t}\text{$\Gamma$}_{0}\lambda-128A_{r}\pi\text{$\beta$}_{t}\text{$\Gamma$%
}_{0}\lambda-16\text{$\alpha$}_{r}\text{$\Gamma$}_{0}\lambda^{2}%
+20B_{t}\text{$\alpha$}_{r}\text{$\Gamma$}_{0}\lambda^{2}\right.  \nonumber\\
& \left.  -20\text{$\beta$}_{t}\text{$\Gamma$}_{0}\lambda^{2}-20A_{r}%
\text{$\beta$}_{t}\text{$\Gamma$}_{0}\lambda^{2}.\right.
\end{align}
As we can see, Eq.(\ref{q14}) shows that all coefficients must be null in
order to satisfy the identity. This restriction gives us the constraint%
\begin{equation}
\alpha_{r}=-\frac{4\beta_{t}\lambda}{36\pi+11\lambda}.\label{q15}%
\end{equation}

\begin{figure}[h!]
\begin{center}
  \includegraphics[width=1\columnwidth]{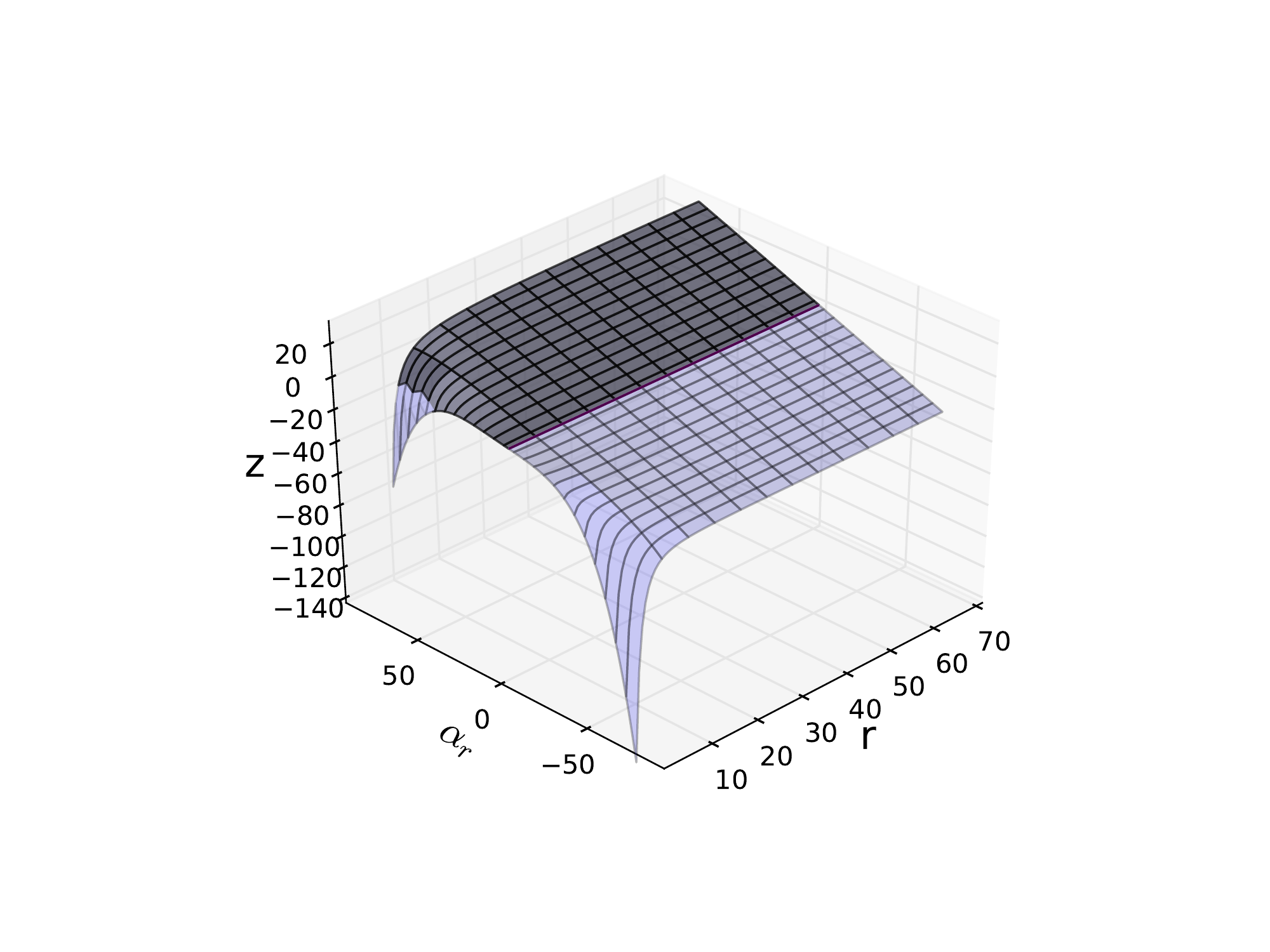}
\end{center}
\caption{Validity of WEC. $z=\rho+\alpha_{r}+A_{r}\rho$ and $\lambda=-1/2$. The dark region is for positive values of $z$ and the clear region, for negative values of $z$.}
\label{alphafig}
\end{figure}

Now we can analyze the WEC for the SWHs in $f(R,T)$ gravity. It reads

\begin{align}
  \rho+p_{r}\geq0,\label{wec1}\\
  \rho+p_{t}\geq0.\label{wec2}
  \end{align}  
By using \eqref{q1} and making $\lambda=-1/2$, from NEC, the first criteria \eqref{wec1} yields  $\rho+\alpha_{r}+A_{r}\rho\geq0$. In Figure \ref{alphafig} below we present the regions in which such an inequality is satisfied.

\begin{figure}[h!]
\begin{center}
  \includegraphics[width=1\columnwidth]{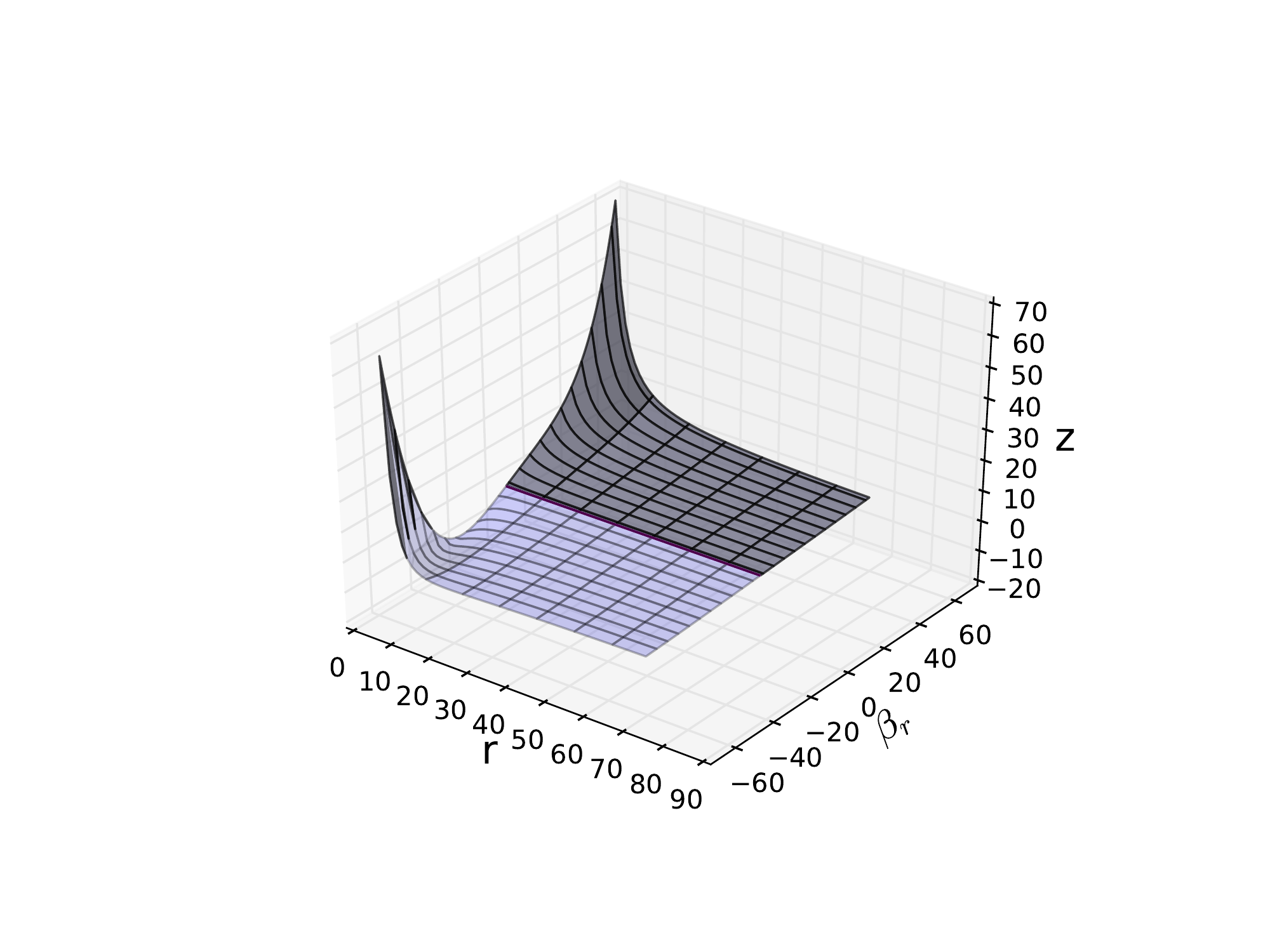}
\end{center}
\label{betafig}
\caption{Validity of WEC. $z=\rho+\beta_{t}+B_{t}\rho$ and $\lambda=-1/2$. The dark region is for positive values of $z$ and the clear region, for negative values of $z$.}
\end{figure}

Now considering the second criteria \eqref{wec2}, using \eqref{q2} we obtain $\rho+\beta_{t}+B_{t}\rho\geq0$. In Figure \ref{betafig} we show the validity regions of such a condition.

\section{Discussion}

\label{sec:dis}

We have approached SWHs from the $f(R,T)$ theory of gravity perspective.
Because of the recent development of such a gravitational model, a further
and careful investigation of WHs in this theory was still missing. That is not
the case for many other alternatives theories of gravity. Besides the
references in the Introduction, WHs have already been studied in Chaplygin gas
models \cite{lobo/2006}, Brans-Dicke theory of gravitation
\cite{nandi/1998,agnese/1995}, Palatini $f(R)$ gravity
\cite{bambi/2016,capozziello/2012} and $f(\mathcal{T})$ gravity (with
$\mathcal{T}$ the torsion scalar) \cite{sharif/2013}, among others. Even WHs
in the presence of a cosmological constant have already been constructed
\cite{richarte/2013b,lemos/2003}.

In a sense, the work we have presented here can be generalized in two forms:
$1)$ by allowing the WH throat radius to vary with time and $2)$ by assuming
an $f(R,T)$ functional form which describes a non-minimal matter-geometry
coupling, such as $f(R,T)=R+f_{1}(R)f_{2}(T)$, with $f_{1}(R)$ and $f_{2}(T)$
being functions of $R$ and $T$, respectively. Certainly the two proposals
above could be considered in the same model, to yield more general solutions.
The consideration of the proposal $\# 2$ in a further work can be compared
with some predictions already obtained for matter-geometry coupling models
others than $f(R,T)$ gravity \cite{garcia/2011,garcia/2010}.

For now, let us analyse our results. We have obtained for the redshift
function a solution like $\varphi(r)\sim\ln(1/r)$. The same $r$-dependence for
$\varphi(r)$ was firstly obtained by Morris and Thorne in \cite{morris/1988}.
The same reference also presents for a SWH with exotic matter limited to the
throat vicinity a density $\rho\sim1/r^{2}$ as our solution (\ref{q13}).
Withal, the same originally proposed SWH presents as a solution for the shape
function $b(r)\sim r^{3}$ for $R_{S}\leq r\leq R_{S}+\Delta R$, with $R_{S}$
being the Schwarzchild radius of the WH and $\Delta R=R_{S}/100$
\cite{morris/1988}. The same proportionality can be appreciated in our
solution (\ref{q5}).

Some few years after the discovery that a cosmological constant is capable of
describing the low brightness of Type Ia Supernovae
\cite{riess/1998,perlmutter/1999}, J.P.S. Lemos et al. have investigated
Morris-Thorne WHs in the presence of a cosmological constant \cite{lemos/2003}%
. For $\varphi(r)$ and $b(r)$, the authors have also obtained the radial
proportionality we have here obtained, i.e., $\ln(1/r)$ and $r^{3}$,
respectively. The argumentation is not quite surprising for the following
reason. The $f(R,T)$ gravity authors have shown that the
$f(R,T)=R+2\lambda T$ cosmological model can be interpreted as a cosmological
model with an effective cosmological constant \cite{harko/2011}. In this way,
by interpreting our model through this approach, it is, somehow, expected to
obtain the same features as in \cite{lemos/2003}.

Here it is worth emphasizing that departing from many works in the literature
(check \cite{azizi/2013,zubair/2016b} for $f(R,T)$ gravity applications and
\cite{jawad/2015,jamil/2013,jamil/2009}, among many others, for other
gravitational models), our solutions have been obtained by considering a
varying redshift function, i.e., $\varphi^{\prime}(r)\neq0$.

Not only our solutions were obtained for a variable redshift function, they
have also been obtained with no assumptions for $b(r)$, departing from
\cite{myrzakulov/2016,zubair/2016b,jamil/2013,yue/2011,rahaman/2006}, among
many others.

Anyhow, our solutions for $\varphi(r)$ and $b(r)$ respect the requirements presented in Section \ref{sec:swhm}.

About the energy conditions application, the NEC and WEC are respected for a wide range of values of the quantities involved.  Such a respectability makes unnecessary to invoke the existence of exotic matter for the WH we have constructed.

Finally, we would like to remark that our calculations have been done
through a direct and exact construction. We have obtained a complete set of
analytical solutions, departing, for instance, from \cite{zubair/2016b}, in which numerical solutions were obtained. An important consequence of our analytical approach is the fact that the same method can be applied in similar scenarios for another alternative gravity theories.

\bigskip

\begin{acknowledgments}
PHRSM would like to thank S\~ao Paulo Research Foundation (FAPESP), grant
2015/08476-0, for financial support. RACC thanks CAPES for financial support. RVL thanks CNPq (Conselho Nacional de Desenvolvimeno Ciet\'ifico e Tecnol\'ogico).
\end{acknowledgments}

\end{document}